\documentclass[10pt,a4paper]{article}
\date{}
\usepackage{float}
\usepackage{graphicx}
\usepackage[english]{babel}
\usepackage{amssymb}

\title{Vacuum Persistence Amplitude and Gravitational Potential in $f(R)$ Gravity}

\author{A. Jahan and S. Heydarnezhad\\
Research Institute for Astronomy and Astrophysics of Maragha (RIAAM)\\
Maragha, P. O. Box: 55134-441, Iran\\
jahan@riaam.ac.ir}
\begin{document}
\maketitle
\begin{abstract}
The source theory provides a straightforward way to obtain the Newton's potential upon establishing the vacuum-to-vacuum transition amplitude in quantized Einstein theory of gravity. Here, we use the same method to derive the gravitational potential of two static point masses in $f(R)=R+aR^2$ gravity.\\
keywords: source theory; vacuum persistence amplitude; $f(R)$-gravity; gravitational potential.
\end{abstract}
\maketitle

\section{Introduction}
Source theory, first invented and developed by Julian Schwinger, is an alternative formulation of quantum field theory which avoids some of the well-known difficulties of the operator formalism. This elegant, powerful yet economic formalism has been extensively applied to particle physics [1, 2]. It has been also fruitful in gravitational physics. For example, using the source theory one can derive the Einstein's gravitational radiation formula in an economic and elegant way. Utilizing the same formalism, one easily derives the Newton's potential for the masses mediating graviton [1-3].\\
In a recent study the source theory is used to study quantized modified gravity with Lagrangian density $f(R)=R+aR^2$ where the vacuum persistence amplitude (or vacuum-to-vacuum transition amplitude) is constructed to calculate the average number of emitted gravitons [4]. The so called modified theories of gravity, is an attempt to extend the Einstein's gravitational action to include the terms of higher powers in curvature tensor [5, 6].\\
In this letter, we use the vacuum persistence amplitude of $f(R)=R+aR^2$ model to derive the gravitational energy of two static point-masses. The result is in accordance with the result obtained in [7-9], where the gravitational potential in quantized $f(R)=R+aR^2$ model is derived by calculating the scattering amplitude of two massive scaler fields mediating gravitons. However, as we are going to show, the source theory is a straightforward method to obtain the static potential even within context of modified theories of gravity. Our choice for the signature of Minkowski metric is $(-,+,+,+)$.
\section{Propagator of Linearized Model}
The action for a $f(R)=R+aR^2$ model, interacting with the matter is
\begin{equation}\label{1}
S=\frac{1}{\kappa^2}\int d^4x\sqrt{-g}\big(R+aR^2\big)+\int d^4x\sqrt{-g}g_{\mu\nu}T^{\mu\nu},\qquad\kappa^2= 16\pi G.
\end{equation}
where $g_{\mu\nu}$ denotes the space-time metric and $T_{\mu\nu}$ is the energy-momentum tensor of matter. This action can be linearized by expanding the metric around the Minkowski metric $\eta_{\mu\nu}$ via
\begin{equation}\label{1}
g_{\mu\nu}=\eta_{\mu\nu}+h_{\mu\nu}.
\end{equation}
The fluctuating field $h_{\mu\nu}$ is related to the energy-momentum tensor through
\begin{equation}\label{1}
h^{\mu\nu}(x)=\int d^4x'\Delta^{\mu\nu\alpha\beta}_+(x-x')T_{\alpha\beta}(x').
\end{equation}
On imposing the gauge condition
\begin{equation}\label{1}
\partial_\mu\Big[h^{\mu\nu}-\frac{1}{2}\eta^{\mu\nu}h^\alpha_\alpha-2a\eta^{\mu\nu}\big(\partial_\alpha\partial_\beta h^{\alpha\beta}-\Box h_\alpha^\alpha\big)\Big]=0,
\end{equation}
and after a lengthy calculation, the Fourier transform of propagator is found to be [4]
{\setlength\arraycolsep{2pt}
\begin{eqnarray}
\nonumber
\widetilde{\Delta}^{\mu\nu\alpha\beta}_+(k)&=&\int{d^4x}{\,\Delta}^{\alpha\beta\mu\nu}_+(x)e^{-ik\cdot x} \\\nonumber
&=&\frac{\eta^{\alpha\mu}\eta^{\beta\nu}+\eta^{\alpha\nu}\eta^{\beta\mu}-\eta^{\alpha\beta}\eta^{\mu\nu}}{2k^2}+
\frac{\eta^{\alpha\beta}\eta^{\mu\nu}}{\frac{1}{a}+6k^2}\\\nonumber
&+&(\lambda-1)\frac{k^\mu k^\beta\eta^{\nu\alpha}+k^\mu k^\alpha\eta^{\nu\beta}+k^\nu k^\beta\eta^{\mu\alpha}+k^\nu k^\alpha\eta^{\mu\beta}}{2k^4}\\
&-&4a\frac{k^\alpha k^\beta k^\mu k^\nu}{k^4}.
\end{eqnarray}}
where $\lambda$ in the third line is an arbitrary gauge parameter. It appears when one incorporates the gauge condition (4) into the linearized Lagrangian as a constraint [4].
\section{Vacuum Persistence Amplitude }
In source theory, the vacuum-to-vacuum transition amplitude plays a central role. For the gravitational theories, its variation with respect to the energy-momentum tensor yields the expectation value of the metric  [1-3]
\begin{equation}\label{1}
\frac{1}{i}\frac{\delta}{\delta T_{\mu\nu}(x)}\langle0_+|0_-\rangle=\langle0_+| h_{\mu\nu}(x)|0_-\rangle.
\end{equation}
where $|0_{\pm}\rangle$ stand for the vacuum (no graviton) state. The explicit form of the transition amplitudes is given by [1-3]
\begin{equation}\label{1}
\langle0_+|0_-\rangle=\exp\bigg[4\pi G \,i\int d^4x\int d^4x'\,T_{\alpha\beta}(x)\Delta^{\alpha\beta\mu\nu}_+(x-x')T_{\mu\nu}(x')\bigg].
\end{equation}
Conservation of energy-momentum tensor, i.e. $\partial_\alpha T^{\alpha\beta}=0$ in momentum space leads to $k_\alpha \widetilde{T}^{\alpha\beta}(k)=0$. Thus, inserting (5) in (7) and using $k_\alpha \widetilde{T}^{\alpha\beta}(k)=0$, yields [4]
{\setlength\arraycolsep{2pt}
\begin{eqnarray}
\nonumber
\langle0_+|0_-\rangle&=&\exp\bigg[4\pi G \,i\int d^4x\int d^4x'\,T_{\alpha\beta}(x)\Big(\eta^{\alpha\mu}\eta^{\beta\nu}-\frac{1}{2}\eta^{\nu\mu}\eta^{\beta\alpha}\Big)D_+(x-x')T_{\mu\nu}(x')\bigg]\\
&\times&\exp\bigg[4\pi G \,i\int d^4x\int d^4x'\,\frac{T_{\alpha}^\alpha(x)}{\sqrt 6}D_+\Big(x-x';\frac{1}{6a}\Big)\frac{T_{\beta}^{\beta}}{\sqrt 6}(x')\bigg],
\end{eqnarray}}
with the massless and massive propagators defined to be
{\setlength\arraycolsep{2pt}
\begin{eqnarray}
D_+(x-x')&=&\int\frac{d^4k}{(2\pi)^4}\frac{e^{ik\cdot(x-x')}}{k^2-i\epsilon},\\
D_+\Big(x-x';\frac{1}{6a}\Big)&=&\int\frac{d^4k}{(2\pi)^4}\frac{e^{ik\cdot(x-x')}}{k^2+\frac{1}{6a}-i\epsilon}.
\end{eqnarray}}
We note that the third and forth terms of (4) make no contribution to (8) due to the conservation of energy-momentum tensor. The first of exponent in (8) is reminiscence of the Einstein gravity [1-3], but the second term appears due to $aR^2$ term of modified action (1).
\section{Gravitational Potential}
In quantum filed theory one can obtain the gravitational potential by taking the Fourier transform of the scattering amplitude for the particles mediating the graviton
\begin{equation}\label{1}
V=-\frac{1}{4m_1m_2}\int\frac{d^3k}{(2\pi)^3}\mathcal{M}_{NR}(k)e^{-i\scriptsize{\textbf{k}\cdot\textbf{x}}}.
\end{equation}
Here $\mathcal{M}_{NR}(k)$ is the scattering amplitude in non-relativistic limit. In source theory setup, the potential $V$ is accessible from the vacuum persistence amplitude via the relation
\begin{equation}
\langle0_+|0_-\rangle=\lim_{T\rightarrow\infty}e^{-iVT}.
\end{equation}
where $T$ denotes the time interval. So, from (8) and (12) one obtains
{\setlength\arraycolsep{2pt}
\begin{eqnarray}
\nonumber
V&=&\lim_{T\rightarrow\infty}-\frac{1}{T}\bigg[4\pi G\int d^4x\int d^4x'\,T_{\alpha\beta}(x)\Big(\eta^{\alpha\mu}\eta^{\beta\nu}-\frac{1}{2}\eta^{\nu\mu}\eta^{\beta\alpha}\Big)D_+(x-x')T_{\mu\nu}(x')\\
&+&4\pi G \int d^4x\int d^4x'\,\frac{T_{\alpha}^\alpha(x)}{\sqrt 6}D_+\Big(x-x';\frac{1}{6a}\Big)\frac{T_{\beta}^{\beta}}{\sqrt 6}(x')\bigg].
\end{eqnarray}}
For two static sources with different masses $m_1$ and $m_2$ located at $\textbf{x}_1$ and $\textbf{x}_2$, the energy-momentum tensor reads
\begin{equation}\label{1}
T^{\mu\nu}(\textbf{x})=\eta^{\mu 0}\eta^{\nu 0}\big[m_1\delta(\textbf{x}-\textbf{x}_1)+m_2\delta(\textbf{x}-\textbf{x}_2)\big].
\end{equation}
Plugging this relation into (13), using (9) and (10) and taking into account
{\setlength\arraycolsep{2pt}
\begin{eqnarray}\label{1}
\int_{-\infty} ^\infty dt\int_{-\infty} ^\infty dt'e^{-i\omega(t-t')}&=&\lim_{T\rightarrow\infty}2\pi T\delta(\omega),\\
\int\frac{d^3k}{(2\pi)^3}\frac{e^{i\scriptsize{\textbf{k}\cdot(\textbf{x}_2-\textbf{x}_1)}}}{\textbf{k}^2+m^2}
&=&\frac{1}{4\pi}\frac{e^{-m\scriptsize{|\textbf{x}_2-\textbf{x}_1|}}}{|\textbf{x}_2-\textbf{x}_1|},
\end{eqnarray}}
gives rise to the gravitational potential as
{\setlength\arraycolsep{2pt}
\begin{eqnarray}
\nonumber
V(r)&=&-4\pi G\int\frac{d^3k}{(2\pi)^3}e^{i\scriptsize{\textbf{k}\cdot(\textbf{x}_2-\textbf{x}_1)}}\bigg(\frac{1}{\textbf{k}^2}
+\frac{1}{3}\frac{1}{\textbf{k}^2+\frac{1}{6a}}\bigg)+\textrm{self-energies}\\
&=&-Gm_1m_2\bigg(\frac{1}{r}+\frac{1}{3}\frac{e^{-Mr}}{r}\bigg)+\textrm{self-energies}.
\end{eqnarray}}
where $r=|\textbf{x}_2-\textbf{x}_1|$ is the distance between the masses and $M^2=\frac{1}{6a}$. Here, the first term is the Newton potential. The second Yukawa-like term arises due to the modification of Einstein-Hilbert action. A similar result is reported in [7-9] where the static potential is obtained upon invoking the formula (11) by calculating the fourier transform of the scattering matrix. However, as we demonstrated in this letter, the source theory method is a rather straightforward way to obtain the gravitational potential.
\section{Conclusion}
We used the source theory to derive the gravitational potential of two static point-masses in modified gravity. The result is in agreement with the result derived using the standard field theory calculations.
\section{Acknowledgment}
The work of S. Heydarnezhad has been supported financially by Research Institute for Astronomy \& Astrophysics of Maragha (RIAAM) under project No.1/5440-37.

\end{document}